# A "configurational-entropy-loss" law for the non-Arrhenius relaxation in disordered systems

Yi-zhen Wang[1,2], X. Frank Zhang[2,*] and Jin-xiu Zhang[1,*]

[1]State Key Laboratory of Optoelectronic Materials and Technologies, and School of Physics and Engineering, Sun Yat-sen University, Guangzhou, 510275, China;

[2]Bioengineering Program and Department of Mechanical Engineering and Mechanics, Lehigh University, 19 Memorial Drive West, Bethlehem, Pennsylvania 18015, USA.

Based on Nowick's self-induced ordering theory, we develop a new configurational-entropy relation to describe the non-Arrhenius temperature($T$)-dependent relaxation in disordered systems. Both the configurational-entropy loss and the coupling interaction among relaxing units (RUs) are explicitly introduced in this relation; thus, it offers a novel connection between kinetics and thermodynamics that is different from the Adam-Gibbs (A-G) entropy relation, and it generalizes several well-known currently used relations. The present relation can provide direct and more accurate estimates of (i) the intrinsic activation enthalpy, (ii) the $T$-evolvement of the systematic configurational entropy loss and (iii) the self-induced ordering temperature $T_c$, which characterizes the coupling interaction among RUs. Application of the theory to experimental relaxation-time data for typical organic liquids demonstrates its validity.

[*] To whom correspondence may be addressed. E-mail: xiz310@lehigh.edu or stszjx@mail.sysu.edu.cn

Relaxation is a universal phenomenon in nature. It refers to the process of matter changing from one equilibrium state to another under an externally applied (e.g., mechanical, electrical or thermal, etc.) field, and it is characterized by an intrinsic parameter, such as the Maxwell "alpha" relaxation time $\tau$, the viscosity $\eta$, or the self-diffusion constant $D_s$. If the relaxation dynamics are dominated by barriers that are to be overcome by thermal fluctuations, the Arrhenius law with the $T$-independent activation enthalpy $\Delta H$ (i.e.[1], $\tau \propto exp(\Delta H/kT)$, where $k$ is the Boltzmann's constant) is often expected to apply. However, this phenomenological law is invalid for most disordered systems such as glass-forming liquids[2-5], proteins[6-8], disordered ferromagnets and ferroelectrics[4,9]. Upon cooling towards the glass transition, many disordered systems exhibit the non-Arrhenius $T$-dependent relaxation behavior. If one insists on using the Arrhenius relation[1], the activation enthalpy must depend on temperature, i.e., $\Delta H(T)$ defined by[4,10] $\tau=\tau_0 exp[\Delta H(T)/kT]$. This relation is easily accepted because the activation enthalpy on temperature can have a slow dependence due, for example, to thermal expansion of the lattice[10]; furthermore, no theoretical basis claims that $\Delta H$ must be strictly independent of $T$. However, it is difficult to assign a recognized physical meaning to the activation energies that arise with decreasing temperature. It is also challenging to derive an explicit and analytical form of $\Delta H(T)$ with a physical meaning at the microscopic level, despite its recognition by a number of research groups as an entropy-loss process due to the interaction among RUs[2,11-13].

In the past centuries many attempts have been made to address this puzzle. The first attempt introduces $\Delta H(T) \equiv \Delta H_0 T/(T-T_0)$, leading to the empirical Vogel-Fulcher-Tammann (VFT) law[14] $\tau=\tau_0 exp[\Delta H_0/k(T-T_0)]$. The VFT law has been experimentally observed in a variety of disordered systems, but its derivation at the micro- and mesoscopic levels and the interpretation of its fitting parameters ($\tau_0$, $\Delta H_0$ and $T_0$) remain insufficient[3,4,15]. The second attempt was the Dienes expression[16]. First formulated by Dienes[16] by considering the structural change of the system via short-range order formalism, the Dienes expression takes the form[16] $\tau=A_D exp[\Delta H_D/kT+B_D/k(T-T_D)]$. The Dienes expression is often considered undesirable because it contains an additional adjustable parameter[17]. The third attempt uses various free-volume models[3, 4, 17] based on the idea that molecules need "free" volume to rearrange. Free-volume models, with additional assumptions[3, 4, 17], can reproduce both the VFT and Dienes relations, but they have the problem:[3, 4, 17] it is impossible to define free volume rigorously, and the relaxation time should not be a function of only density. Another well-known attempt is the *A-G* entropy model[18]. By assuming that relaxation involves the cooperative/coupling rearrangement of RUs, it predicts that[18] $\tau=\tilde{A}exp[C/TS_{conf}(T)]$, which provides an attractive physical scenario to describe the contribution of the configurational entropy ($S_{conf}$) to the non-Arrhenius relaxation process. Despite the domination of the *A-G* entropy theory in this field for many years, in recent decades scientists have realized that it is not a panacea[3, 4, 9, 15]. Many other approaches[3, 4, 9, 17] have also been proposed.

However, the non-Arrhenius relaxation in disordered systems remains incompletely understood and leaves several unanswered questions:[4, 9, 10, 17, 19] the cause of the non-Arrhenius relaxation, the link between its kinetic and thermodynamic properties and the roles of the configurational entropy and the coupling interaction among RUs. Here we present a new configurational entropy theory that provides a comprehensive phenomenological physical picture of the non-Arrhenius relaxation in disordered systems, where the influences of the configurational entropy and the coupling interaction among RUs are considered. The theory successfully accounts for the dielectric relaxation in typical organic liquids.

The theory that we shall explore is based on Nowick's stress-induced ordering (SIO) theory[10, 20]. We consider a generic disordered system that possesses a conserved total concentration $C_0$ of randomly arranged elastic dipoles of the λ-tensor[20] (or electric dipole impurities of the electric dipole moment[10] $\mu$) with two crystallographically equivalent orientations (denoted by $p=1$ and *2*) that have the same energy in the absence of an external field. Because the total dipole concentration $C_0$ is conserved (that is $\sum_{p=1}^{2} C_p = C_0$, where $C_p$ denotes the concentration of dipoles in the *p* orientation), the system can be described by an independent order parameter *c*, which is defined as the deviation of $C_1$ from its average value $C_0/2$, i.e., $c \equiv C_1 - C_0/2$. The configurational entropy of the system, which can be measured statistically from the number of ways of dividing $C_0$ objects into two groups with $C_1$ in the first group and $C_2$ in the second, is expressed in terms of the variable $c$ as [20]:

$$S_{conf} = \frac{1}{2}kN_A C_0 \left\{ \ln 2 - [(1+\frac{2c}{C_0})\ln(1+\frac{2c}{C_0}) + (1-\frac{2c}{C_0})\ln(1-\frac{2c}{C_0})] \right\}, \qquad [1]$$

or, by expanding for small values of $c$, $S_{conf} \approx (1/2)kN_A C_0 \ln 2 - (2kN_A/C_0)c^2$, where $N_A$ is the Avogadro's number. Applying an external field (e.g.[†], a uniaxial stress $\sigma$) to the system would cause the free energies of two crystallographic orientations of the dipole to split, as illustrated in Fig. 1. This setup provides the basis for the redistribution of dipoles in the system; therefore, a relaxation process ensues.

***Thermodynamic description:*** As in the SIO theory[20], we model the ensued relaxation process as a purely dipole-diffusional transformation defined by the generalized Gibbs free energy function:[‡]

$$G(\sigma, c, T) = G(0,0,T) - \frac{1}{2}V_0 J_U \sigma^2 - V_0(\delta\lambda)\sigma c - V_0 L(T-T_{ref})\sigma - \frac{1}{2}N_A bc^2 - TS_{conf}, \qquad [2]$$

where $V_0$ is the molar volume, $J_U$ is the instantaneous compliance of the system, $\delta\lambda \equiv \lambda^{(1)} - \lambda^{(2)}$ and $\lambda^{(p)}$ ($p=1, 2$) is the tensile component of the $\lambda$-tensor[10, 20] along the stress axis of the elastic dipole of type $p$ (or the component of the electric dipole moment along the electric field), and $b$ is the attractive interaction coefficient per molecule (with a positive value in the present calculation as Nowick did[20]). In Eq. 2, the first term characterizes the free energy of the equilibrium state when $\sigma=0$. The second and third terms describe the contribution due to the applied $\sigma$. The fourth term is the thermal expansion due to the temperature change of the system relative to an arbitrary reference temperature[20] $T_{ref}$. The last two terms represent the contributions from the coupling interaction among RUs (i.e., dipoles) and the configurational entropy, respectively.

Since the system (at certain $\sigma$ and $T$) achieves thermal equilibrium when the Gibbs free energy functional is minimum (that is, $\partial G/\partial c=0$)[10, 20], using Eq. 2 and the quadratic approximation for $S_{conf}$ (for simplification), one can obtain the equilibrium value $\bar{c}$ given by[20]

$$\bar{c} = \frac{C_0 v_0 (\delta\lambda)\sigma}{4k(T-T_c)}, \qquad [3]$$

where $v_0 \equiv V_0/N_A$ and $kT_c \equiv bC_0/4$. Here, $T_c$ is the "self-induced ordering"[10, 20] critical temperature, which is analogous to $T_0$, $T_D$ (in VFT[14] and Dienes[16] relations) and the Kauzmann temperature[13] $T_K$. In the absence of interaction (i.e., $b=0$), $T_c=0$, and $\bar{c}$ is inversely proportional to $T$. Eq. 3 indicates the existence of the thermodynamic equilibrium state with an order singularity at $T_c$ due to the interaction among RUs.

---

[†]To simplify the calculation, the external field applied here is given by the uniaxial stress, σ, oriented so as to distinguish two dipole orientations, as Nowick did in (19).
[‡]This treatment is analogous to that used by Nowick (19) to describe the order-singularity-like relaxation strength in crystals due to the stress-induced ordering behavior. There is no pure $c$-dependent term in Eq. 2 because for σ=0, the value $c$=0 corresponds to a state of equilibrium (19).

*Kinetic description:* To complete the model, the kinetics of the ensued relaxation process must also be defined. As illustrated in Fig. 1, a free energy barrier ($\Delta F^0$) must be surmounted when a dipole jumps from one orientation to another, and the free energy levels of the RUs split by an amount $\Delta g$ after the stress $\sigma$ is applied. The quantity $\Delta g$, which is defined as the difference in free energy when a specific dipole is converted from a type-2 orientation to a type-1 orientation, is given by[20]

$$\Delta g \equiv \frac{1}{N_A}\frac{\partial(G+TS_{conf})}{\partial c} = -v_0(\delta\lambda)\sigma - bc. \qquad [4]$$

Thus, the probability frequencies of reorientation of a dipole in the presence of $\sigma$ (defined as $v_+$ if the dipole goes from 1 to 2, and $v_-$ from 2 to 1) may be expressed as[10]

$$v_\pm = \tau_0^{-1}\exp(-\frac{\Delta F^0 \pm |\Delta g|/2}{kT}), \qquad [5]$$

where $\tau_0^{-1}$ is an appropriate average lattice vibration frequency, and $\Delta F^0$ is given by[10] $\Delta F^0 = \Delta H^0 - T\Delta S_v$ in terms of an activation enthalpy $\Delta H^0$ and a vibrational activation entropy $\Delta S_v$. The kinetic equation to describe the rates of change of the number of dipoles in the orientation $p=1,2$ can then be described in terms of $c$ as follows:[10]

$$dc/dt = -\tau^{-1}[c-\bar{c}_k], \qquad [6a]$$

where $\bar{c}_k = \frac{C_0}{2}(\frac{v_- - v_+}{v_- + v_+})$  [6b]

and $\tau^{-1} = v_- + v_+$.  [6c]

Eq. 6 describes the kinetics of RUs relaxing toward the kinetic equilibrium state $\bar{c}_k$ (where *dc/dt=0*) with the average rate ($\tau^{-1}$). From Eqs. 4-6, we note that both $\bar{c}_k$ and $\tau^{-1}$ are complex functions of the splitting energy at $\bar{c}_k$ (i.e., $\Delta g(\bar{c}_k)$), which depends on both the applied stress and the interaction term "$-b\bar{c}_k$". The "$-b\bar{c}_k$" term can reduce the free energy to generate order[20]. However, this interaction term is not introduced in the general theory of kinetics of molecular relaxation[10], which hence derived the *T*-dependence of $\bar{c}_k$ (as predicted by Eq. 3 for $T_c=0$) and the Arrhenius relation of $\tau^{-1}$. The present model attempts to investigate the case when considering the interaction term, i.e., $-b\bar{c}_k \neq 0$.

As usual[10], if we approximate Eq.5 (i.e., $v_+ \approx \tau_0^{-1} exp(-\Delta F^0/kT)(1-|\Delta g|/2kT)$ and $v_- \approx \tau_0^{-1} exp(-\Delta F^0/kT)(1+|\Delta g|/2kT)$, which are obtained under $|\Delta g|/kT<<1$), it is possible to simplify Eq. 6b to $\bar{c}_k \approx -\Delta g(\bar{c}_k)C_0/4kT$. Thus, replacing $\Delta g(\bar{c}_k)$ in this simplified formula with Eq. 4 (in $\bar{c}_k$) gives $\bar{c}_k \approx \frac{C_0 v_0(\delta\lambda)\sigma}{4k(T-T_c)}$, which is consistent with Eq. 3. If the interaction term "$-b\bar{c}_k$" is neglected so that $\Delta g(\bar{c}_k) \approx -v_0(\delta\lambda)\sigma$, the simplified formula shows that $\bar{c}_k$ is also inversely proportional to *T*. The above analysis explicitly illustrates that the order-singularity equilibrium state (described by Eq. 3) exists in both the thermodynamic and the kinetic representations, and the involved coupling interaction manifests its entire existence in $T_c$.

***Definition of the configurational-entropy-loss order parameter* $\Delta s^*_{conf}$:** How is the average relaxation rate ($\tau^{-1}$, described in Eq. 6c) expressed when we consider the interaction term "-$b\bar{c}_k$"? The above discussion indicates that the rate should be more complex than an Arrhenius-type

description$^§$. To clarify this issue, we first introduce the relevant order parameter $\Delta s^*_{conf}$.

Because $\partial G/\partial c |_{c=\overline{c}_k}=0$, from Eq. 4 we can have

$$\Delta g(\overline{c}_k) = T\frac{1}{N_A}\frac{\partial(S_{conf})}{\partial c}|_{c=\overline{c}_k} \equiv -T\Delta s^*_{conf}, \qquad [7]$$

which shows that $\Delta g(\overline{c}_k)$ is related to the configurational entropy change of the system at $\overline{c}_k$. By substituting $\Delta g(\overline{c}_k) = -T\Delta s^*_{conf}$ into the approximating relation of Eq.6b (i.e., $\overline{c}_k \approx -\Delta g(\overline{c}_k)C_0/4kT$) and using Eq. 3, we find that the quantity $\Delta s^*_{conf}$, defined by $\Delta s^*_{conf} \equiv -N_A^{-1}(\partial S_{conf}/\partial c)|_{c=\overline{c}_k}$, satisfies

$$\Delta s^*_{conf} = \frac{4k\overline{c}_k}{C_0} = \frac{v_0(\delta\lambda)\sigma}{T-T_c}. \qquad [8]$$

Moreover, substituting $\Delta s^*_{conf}=4k\overline{c}_k/C_0$ into the quadratic approximation in Eq. 1 (in $\overline{c}_k$) can give the relationship between $S_{conf}$ and $\Delta s^*_{conf}$

$$S_{conf} \approx S_{conf\_0} - (C_0 N_A/8k)(\Delta s^*_{conf})^2, \qquad [9]$$

where $S_{conf\_0} \equiv S_{conf}(\Delta s^*_{conf}=0)$ represents the case when $\Delta s^*_{conf}=0$. Eqs. 8 and 9 state that the order parameter $\Delta s^*_{conf}$ characterizes the structural order degree of the system at $\overline{c}_k$ and the configurational entropy change due to the coupling interaction among RUs under the external relaxation-inducing stress.

*A new configurational-entropy-loss law:* By substituting Eq. 5 into Eq. 6c, we find that the relaxation time can be written as[**] $\tau=\alpha\tau_0 exp[(\Delta F_0-\Delta g(\overline{c}_k))/kT]$, where the interaction term is explicitly introduced. Substitution of $\Delta g(\overline{c}_k)$ by $-T\Delta s^*_{conf}$ then yields:

---

$^§$Nowick (9, 19) argued that it might be controlled by atomic movements governed by an Arrhenius-type equation. Such Arrhenius-type description was usually obtained by neglecting the interaction term. That is (9), the substitution of two approximate forms of Eq. 5, $v_+\approx\tau_0^{-1}exp(-\Delta F^0/kT)(1-|\Delta g|/2kT)$ and $v_-\approx\tau_0^{-1}exp(-\Delta F_0/kT)(1+|\Delta g|/2kT)$, into Eq. 6c leads to $\tau^{-1}\approx 2\tau_0^{-1}exp(-\Delta F^0/kT)$. Such treatment completely covers the interaction term "$-b\overline{c}_k$" in $\tau^{-1}$.

[**] Note that, the use of $\tau=\alpha\tau_0 exp[(\Delta F_0-\Delta g(\overline{c}_k))/kT]$ (rather than $\tau\sim\tau_0 exp[(\Delta F_0+\Delta g(\overline{c}_k))/kT]$) in the present calculation follows from the slowdown property of the disordered system with decreasing temperature; and that the $T$ dependence of the near-unity coefficient $\alpha \equiv [1+exp(-\Delta g/2kT)]^{-1}$ is negligible compared to the exponential term.

$$\tau = A\exp(\frac{\Delta H^0 + T\Delta s^*_{conf}/2}{kT}),  \qquad [10]$$

where $A \equiv \alpha\tau_0 exp(\Delta S_v/k)$ is considered independent of temperature. Eq.10 exhibits the dependence of the relaxation time on the configurational entropy loss of the system due to the coupling interaction among RUs. This relation is different from the A-G entropy law[18] and exhibits the functional form similar to the previously reported excess-entropy scaling law[21, 22].

Insertion of Eq. 8 into Eq. 10 can give a more practical form:

$$\tau = A\exp(\frac{\Delta H^0}{kT} + \frac{B'}{k(T-T_c)}),  \qquad [11]$$

where $2B' \equiv v_0(\delta\lambda)\sigma$ represents[20] the relaxation energy induced by $\sigma$. Eq. 11 resembles the Dienes expression[16], but its constants show different and well defined physical meanings. When the coupling interaction among RUs is so strong that the term $B'/k(T-T_c)$, which represents the configurational entropy change, is much larger than the activation enthalpy term, ($\Delta H^0/kT$), Eq. 11 leads to the VFT law[14]; furthermore, when the concentration of RUs ($C_0$) is so small that the unit interactions do not occur, $T_c=0$, giving rise to the Arrhenius law[1]. Therefore, Eq.10 generalizes the Arrhenius equation [1], the VFT law[14] and the Dienes expression[16].

***Comparison with experiment:*** The entire field of disordered systems that exhibit the non-Arrhenius relaxation is beyond the scope of this paper. To verify our theory, we apply Eq. 11 in conjunction with Eq.10 to analyze the previously reported[15, 23-26] dielectric relaxation data for 7 types of typical organic liquids (shown in inset of Fig. 2), and to estimate their intrinsic *T*-independent activation enthalpy, $T_c$ (characterizing the attractive coupling interaction among RUs), and the *T*-evolvement of the configurational-entropy loss. We chose these dielectric relaxation data because dielectric relaxation measurements give the most precise relaxation-time data[15] (much more accurate than the data from other relaxation processes or from viscosity measurements) and because organic liquids are believed to have the best dielectric data for disordered systems[3, 4, 15]. To quantify how well Eq. 11 fits the data, we compare its fitting with the VFT equation[14], which is considered the most popular fitting function. All fitting analysis in this work is by the least-square method[15, 17, 27]; the data-selecting and subsequent fitting procedures are automated using the ORIGIN software[27, 28].

The inset in Fig. 2 shows all of the analyzed data and the best-fit curves by Eq. 11. All investigated disordered systems exhibit the characteristic non-Arrhenius behavior, and Eq. 11 fits the data well. The best-fit parameters for both Eq. 11 and the VFT fits are listed in Table 1 with the reduced *ChiSqr* (or *Chi^2/DOF*) deviations[27, 28] (that is, $\sigma^2$ and $\sigma_{VFT}^2$) given by ORIGIN[27, 28]. The two parameters, $\sigma^2$ and $\sigma_{VFT}^2$, characterize the deviations from the fits by Eq.11 and the VFT equation to original experimental data, respectively. Despite little visible difference between the Eq. 11 fitting and the VFT fitting (shown in inset in Fig. 2), Table 1 shows that Eq. 11 performs much better than VFT for all liquids investigated: the Eq. 11 fitting yields not only smaller values of $\sigma^2$ but also more reasonable best-fit parameters with better-defined physical meaning than the VFT fit. The values of the prefactor *A* are in (or just slightly outside) the range of $10^{-15} \sim 10^{-13}$ second, coinciding with the natural oscillation frequency of atoms in matters[17]. The $\Delta H^0$ values are consistent with previous experimental results[2, 11, 15, 16, 23-26], representing the intrinsic activation enthalpy of the system in the ideal completely-disordering case, where the intermolecular interaction is negligible. The *B′* values, which are much smaller than corresponding $\Delta H^0$ values, represent the relaxation-triggered energy[20] due to the external field. The $T_c$ values, which are smaller than their corresponding $T_g$ but larger than $T_0$, characterize the different average-meaning attractive interaction (among RUs) in the investigated systems. In contrast, the values of the parameters ($\tau_0$, $\Delta H_0$ and $T_0$) obtained from the VFT fits lack physical meaning[3, 4, 15], and the activation enthalpy values, denoted by $\Delta H_0$, are obviously under-predicted.

Subsequently, from Eq. 11 and Eq. 10 (or Eq. 8), using the *B′* and $T_c$ values provided in Table 1, we can directly calculate the corresponding *T*-dependences of the configurational-entropy loss (i.e., $\Delta s^*_{conf}(T)$), as shown in Fig. 2. Fig. 2 illustrates that the non-Arrhenius kinetic slowing-down of all investigated liquids accompanies the development of the configurational entropy loss and the structural order degree of the system. As the temperature ($T > T_g$) approaches $T_g$, the systematic configurational-entropy loss increases, and the relaxing system becomes more ordered. This $\Delta s^*_{conf}(T)$ behavior is similar to the *T*-evolution of the static structural order for glass-forming liquids recently observed in simulations[19], suggesting that dynamic heterogeneity[3, 19, 29] and fragility[2-4] may develop in the investigated disordered systems. Analogous results for glass-forming liquids, exhibited by the *T*-dependences of the excess entropy $S_{exc}$ (or $S_{conf}$), were also obtained by[11] combining the A-G and VFT equations or by using[2] $S_{conf}=\Delta C_p ln T/T_k$ (where $\Delta C_p$ denotes the excess heat capacity). We emphasize that the present method to calculate $\Delta s^*_{conf}(T)$ is unbiased and more promising than the previous approaches[2, 11-13] Because it makes no priori assumptions about special relaxation and the configurational entropy. In contrast, in previous approaches[2, 11-13], $S_{exc}$ or $S_{conf}$ is usually defined arbitrarily and with some ambiguous assumptions (e.g., $S_{exc} \equiv S_{conf} \equiv S_{liquid} - S_{crystal}$ or $\Delta C_p$ is assumed to be constant), and the origin of the VFT equation is still controversial[3, 4, 15].

In summary, based on relaxation thermodynamics and kinetics in the generic disordered dipole system, we have developed a new configurational entropy theory to explain the non-Arrhenius relaxation in disordered systems. A hallmark of the present model is the "configurational-entropy-loss" law (described by Eq. 10 and its practical form of Eq. 11), where both the configurational entropy loss and the coupling interaction among RUs (characterized by $T_c$) are explicitly introduced. The "configurational-entropy-loss" law not only offers a new connection between kinetics and thermodynamics, different from the A-G entropy relation[18], but it is also shown to generalize several well-known relations[1, 16, 18] in current use. Using this law, one can directly and more accurately extract the intrinsic activation enthalpy, $T_c$ and the *T*-evolution of configurational-entropy loss from the dielectric relaxation spectrum. The theory has been applied to interpret accurate dielectric relaxation for typical organic liquids. Furthermore, the "configurational-entropy-loss" law is expected to be applicable to other types of relaxations such as mechanical/dielectric relaxation in ferroelectric relaxors, magnet relaxation in ferromagnetic and paramagnetic materials.

This work was supported by the Chinese National Physics Base grants Nos. J0630320 and J0730313 (to J.Z.) and the American Heart Association grant 11SDG5420008 (to X.F.Z.).

**Figures and captions**

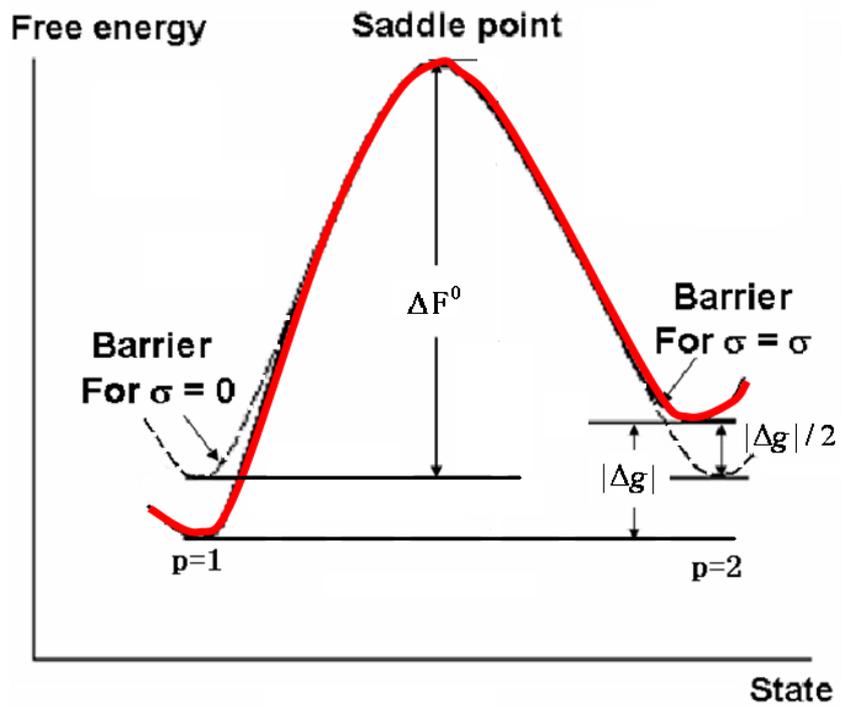

**Fig. 1.** Schematic diagram [10] of the free energy barriers before (dashed lines) and after (solid lines) the free-energy levels split due to the externally applied field σ.

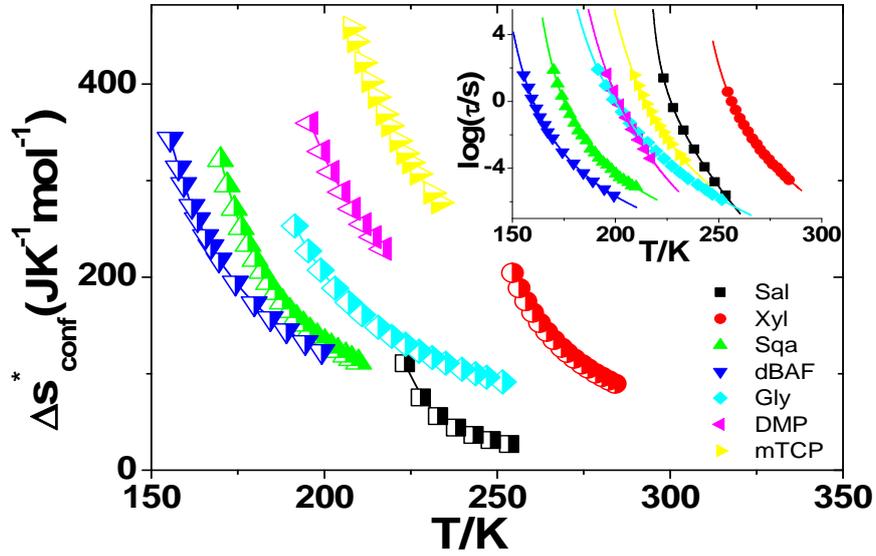

**Fig. 2** Systematic configurational entropy change (characterized by $\Delta s^*_{conf}$) as a function of temperature for typical organic liquids[15, 23-26]. The inset shows the original dielectric relaxation time data (points) and fitting results (lines) by Eq.11. We stress that, $\Delta s^*_{conf}$ is a thermodynamic order variable, characterizing both the structural order degree of the equilibrium system and the configurational entropy loss due to the coupling interaction among RUs; and that it is different from the excess entropy $S_{exc}$, defined by[2,10-12] $S_{exc} \equiv S_{liquid} - S_{crystal}$. Eq.9 describes the relation between $\Delta s^*_{conf}$ and the configurational entropy of the disordered system satisfies.

**Table 1:** Parameters relevant to the Eq. 11 and VFT [14] fits. All data investigated were limited to those above the glass transition temperature ($T_g$), defined by[4] $\tau(T_g)=100s$.

| Liquids (Abr.) | By Eq. 11 | | | | | | By VFT | | | |
|---|---|---|---|---|---|---|---|---|---|---|
| | $\log(A)[s]$ | $\Delta H^0 [10^3 \text{ J/mol}]$ | $B'[10^3 \text{ J/mol}]$ | $T_c[K]$ | $\sigma^2[\%]$ | $T_g[K]$ | $\log(\tau_0)[s]$ | $\Delta H_0[10^3 \text{ J/mol}]$ | $T_o[K]$ | $\sigma_{VFT}^2[\%]$ |
| Sal[23] | -12.31 | 173.61 | 0.57 | 213.49 | 0.17 | 222.12 | -15.91 | 14.46 | 178.01 | 0.64 |
| Xyl[15] | -12.40 | 105.13 | 2.31 | 231.16 | 0.07 | 250.43 | -12.40 | 10.60 | 210.14 | 0.09 |
| Sqa[25] | -14.60 | 26.04 | 3.37 | 148.79 | 0.13 | 169.65 | -10.13 | 5.78 | 141.28 | 0.24 |
| dBAF[24] | -14.13 | 20.25 | 4.14 | 130.93 | 0.19 | 154.62 | -10.71 | 6.75 | 123.83 | 0.30 |
| Gly[26] | -12.67 | 54.01 | 4.24 | 157.35 | 0.07 | 191.14 | -13.93 | 18.32 | 128.56 | 0.13 |
| DMP[15] | -14.61 | 63.65 | 6.65 | 158.66 | 0.06 | 194.89 | -15.94 | 17.36 | 143.09 | 0.12 |
| mTCP[26] | -15.09 | 27.97 | 8.06 | 168.57 | 0.09 | 207.57 | -13.21 | 13.50 | 162.25 | 0.15 |